\documentclass[prb,aps,amsmath,groupedaddress,eqsecnum]{revtex4}

\usepackage{amssymb}
\usepackage{amsmath}
\usepackage{graphicx}
\usepackage{amsbsy}
\usepackage{color}
\newcommand{\bq}{\begin{eqnarray}}
\newcommand{\eq}{\end{eqnarray}}
\newcommand{\bqn}{\begin{eqnarray*}}
\newcommand{\eqn}{\end{eqnarray*}}

\begin{document}
%%%%%%%%%%%%%%%%%%%%%%%%%%%%%%%%%%%%%%%%%%%%%%%%%%%%%%%%%%%%%%%%%%%%%%%%%%%%%%%
%%%%%%%%%%%%%%%%%%%%%%%%%%%%%%%%%%%%%%%%%%%%%%%%%%%%%%%%%%%%%%%%%%%%%%%%%%%%%%%
%%%%%%%%%%%%%%%%%%%%%%%%%%%%%%%%%%%%%%%%%%%%%%%%%%%%%%%%%%%%%%%%%%%%%%%%%%%%%%%
\title{Effects of Polydispersity and Anisotropy in Colloidal and Protein Solutions: an Integral Equation Approach}

\author{Domenico Gazzillo}
\email{gazzillo@unive.it}
\affiliation{Dipartimento di Scienze Molecolari e Nanosistemi, Universit\`a Ca' Foscari Venezia,
Calle Larga S. Marta DD 2137, I-30123 Venezia, Italy}
\author{Achille Giacometti}
\email{achille@unive.it}
\affiliation{Dipartimento di Scienze Molecolari e Nanosistemi, Universit\`a Ca' Foscari Venezia,
Calle Larga S. Marta DD 2137, I-30123 Venezia, Italy}

\date{\today}

\begin{abstract}
Application of integral equation theory to complex fluids is reviewed, with particular emphasis to the effects of polydispersity
and anisotropy on their structural and thermodynamic properties. Both analytical and numerical solutions of integral equations are discussed
within the context of a set of minimal potential models that have been widely used in the literature.
While other popular theoretical tools, such as numerical simulations and density functional theory, are superior for quantitative 
and accurate predictions, we argue that integral equation theory still provides, as in simple fluids, an invaluable technique
that is able to capture the main essential features of a complex system, at a much lower computational cost.
In addition, it can provide a detailed description of the angular dependence in arbitrary frame, unlike numerical simulations
where this information is frequently hampered by insufficient statistics.
Applications to colloidal mixtures, globular proteins and patchy colloids are discussed, within a unified framework.
\end{abstract}

%\pacs{...}
%\keywords{....}

\maketitle
%%%%%%%%%%%%%%%%%%%%%%%%%%%%%%%%%%%%%%%%%%%%%%%%%%%%%%%%%%%%%%%%%%%%%%%%%%%%%%%
\section{Introduction}
\label{sec:introduction}
%%%%%%%%%%%%%%%%%%%%%%%%%%%%%%%%%%%%%%%%%%%%%%%%%%%%%%%%%%%%%%%%%%%%%%%%%%%%%%%
Colloids are dispersions of mesoscopic particles (of the order of 1-500 nm in diameters) suspended in an atomistic fluid. The fact that the typical energy scales 
involved in their interactions are of the order of the thermal energy, constitutes one of the main differences with respect to simple fluids. 
An additional important advantage
arises from the fact that they allow for an almost arbitrary tunability of interactions (unlike their atomistic counterpart fixed by elementary chemistry), 
thus opening the possibility of engineering colloidal systems and synthesizing particles with specific properties.

One of the typical difficulties that one has to cope with in the investigation of colloidal suspensions is associated with their polydispersity. As a matter
of fact, colloids can hardly be synthetized identical to one another, and they often present a range of different chemical properties, such as size or charge.
This fact makes the theoretical analysis of their thermophysical properties rather problematic, a feature that it has been well recognized only
in the last decade. 

When the distribution of these components is wide, one often approximate it with a continous distribution density, and this is referred to as polydispersity.
Physically, the effect of polydispersity is to allow different molecules to pack more closely, thus effectively inhibiting  the formation of an ordered structure.
As a matter of fact, even within the simple hard-sphere description, there are interesting results, such as, for instance, the disappeareance of a fluid-solid transition
for sufficiently large polydispersity. Within simple approximations, (e.g. the Percus-Yevick or the mean-spherical approximations), exact solutions
of the Ornstein-Zernike equations are possible for a number of simple but relevant potential models, including hard-spheres, sticky hard-spheres, 
Coulomb and Yukawa potentials. 

Another important fact that has to be taken into account, when dealing with complex fluids, is the possibility of anisotropic interactions.

Isotropic potentials, i.e. potentials that are independent of the relative orientations of the interacting particles, are often exploited in colloidal physics 
as effective potentials, where the microscopic degrees of freedom are traced out, and thus provide mediated interactions between the larger degrees of freedom. 
On the other hand, there exists a number of systems where this simplifying approximation is certainly not justified, even at the minimal level. 
A paramount example of this is given by globular proteins. A protein is physiologically active when the polypeptide chain forming the primary sequence folds into a 
three-dimensional structure. In spite of the fact that different combinations of the 20 amino acids can, in principle, give rise to a gigantic number of possible 
different three-dimensional structures, only a few thousand folds have been observed, thus suggesting that a number of strong constraints are drastically limiting 
the number of possible three-dimensional architectures. Although the globular shape of many proteins could roughly be approximated to a sphere, 
the presence of different surface groups gives rise to strong anisotropies in the protein-protein interactions.

Another more recent example of anisotropic colloids is given by patchy colloids, which constitutes one recent and very active research topic, in view
of the remarkable applications that they might have in the engineering of future materials. Once again, some of the paradigmatic models used
in simple fluids can be modified and adapted to the new physical situation. Due to the angular dependence included in the potentials, 
some of the techniques to be used can be borrowed from the context of associating (molecular) fluids, although important modifications are required in
view of the discontinuos nature of the angular dependance that typically occurs in patchy colloids and globular proteins. 

This paper aims to review some recent efforts in both directions -- polydispersity and anisotropy -- that have been carried out in the last few years.
While the choice of the topics and the methodologies are mainly dictated by the authors' interests, it will be shown that in both cases
they originated from the landmark studies by Prof. Lesser Blum, that -- directly or indirectly --- influenced all past, present and future work in the field. 

%%%%%%%%%%%%%%%%%%%%%%%%%%%%%%%%%%%%%%%%%%%%%%%%%%%%%%%%%%%%%%%%%%%%%%%%%%%%%%%
\section{From Simple to Complex Fluids}
\label{sec:simple}
%%%%%%%%%%%%%%%%%%%%%%%%%%%%%%%%%%%%%%%%%%%%%%%%%%%%%%%%%%%%%%%%%%%%%%%%%%%%%%%
For spherically symmetric potentials, thermophysical properties of a system of density $\rho$ can be obtained from the Ornstein-Zernike (OZ) equation
\begin{eqnarray}
\label{simple:eq1}
h\left( r\right) =c\left( r\right) +\rho \int d \mathbf{r}^{\prime }~c\left( r^{\prime }\right)
~h\left( |\mathbf{r-r}^{\prime }|\right) 
\end{eqnarray}%
but its solution can be accomplished only in the presence of an additional
approximation (closure) to the exact relation between the direct correlation function (DCF) 
$c\left( r\right) $, the potential $\phi(r)$, and the total correlation function $h\left(r\right)$ that has the general form
\begin{eqnarray}
\label{simple:eq2}
c\left(r\right) &=& \exp{\left[-\beta \phi\left(r\right) + \gamma\left(r\right)+B \left(r\right) \right]}-1 -\gamma\left(r\right)
 \end{eqnarray}
where $\gamma\left( r\right) =h\left( r\right)-c\left( r\right)$. 
Here $\beta\equiv\left(  k_{\mathrm{B}}T\right)  ^{-1}$, $k_{\mathrm{B}}$ is Boltzmann's constant, and $T$ the absolute temperature. 
The closure may be regarded as an approximation to the bridge function $B(r)$, that is known only as an infinite power series in density whose coefficients cannot 
be readily calculated (Hansen and McDonald 2006). All practical closures approximate $B\left(r\right)$ in some way.

As a consequence, different routes
to achieve structural and thermodynamic quantities yield in general different results, the most well known being the compressibility,
the virial (or pressure) and the energy routes (Hansen and McDonald 2006).

Clearly, the goodness of the results depends upon the accuracy of the approximation chosen for the bridge function $B(r)$. Among the most popular
approximations are the Mean Spherical Approximation (MSA), with the DCF being related only to the potential outside the core by
\begin{equation}
\label{simple:eq3}
c\left( r\right) =-\beta \phi\left( r\right) \qquad r\geq \sigma,  
\end{equation}
complemented by the condition of
excluded volume, $h(r)=-1$ inside the hard core of size $\sigma$ (the diameter of a particle); the Percus-Yevick (PY) closure 
\begin{equation}
\label{simpla:eq4}
c\left( r\right) =\left[ e^{-\beta \phi\left( r\right) }-1\right] %
\left[ 1+\gamma\left( r\right) \right] ,  
\end{equation}
 and the Hypernetted Chain (HNC) closure 
\begin{equation}
\label{simple:eq5}
c\left( r\right) =e^{-\beta \phi \left( r\right) +\gamma\left( r\right) }-1-\gamma\left( r\right)  
\end{equation}
In the latter the bridge function is set to zero, with the remarkable advantage of having automatically enforced the equality of the virial
and the energy routes (Morita 1960). An additional improvement is given by the so-called Reference Hypernetted Chain (RHNC), where the bridge function
is taken to be that of the hard-spheres potential with tunable diameter (Lado 1982, 1982a, 1982b). This is selected to reintegrate the energy-virial consistency,
lost with the introduction of the bridge function.

In this respect, here we also mention the self-consistent Ornstein-Zernike approximation (SCOZA), a microscopical liquid-state theory that was proposed by
Stell and collaborators (H\o ye and Stell 1977, Pini \textit{et al} 1998, Kahl \textit{et al} 2002, Sch\"oll-Paschinger \textit{et al} 2005, ), 
that has been shown to be able to predict critical points and lines
of the fluid-fluid coexistence with remarkable accuracy. As in the RHNC case, this is a consequence of the enforcing thermodynamic consistency between the
different routes.

Unlike simple fluids, complex fluids are typically formed by more than one species and hence the above equations have to be extended to mixtures.
As a first example, the case of ionic fluids will be discussed next. 

%%%%%%%%%%%%%%%%%%%%%%%%%%%%%%%%%%%%%%%%%%%%%%%%%%%%%%%%%%%%%%%%%%%%%%%%%%%%%%%
\section{Long-ranged electrostatic interactions: primitive model for
electrolytes}
\label{sec:long}
%%%%%%%%%%%%%%%%%%%%%%%%%%%%%%%%%%%%%%%%%%%%%%%%%%%%%%%%%%%%%%%%%%%%%%%%%%%%%%%
Historically, the first successful theory of ionic solutions was the classic
work by Debye and H\"{u}ckel (1923), based on the solution of the differential
Poisson-Boltzmann equation. In the modern work, based upon the statistical
mechanics of liquids, ionic fluids are studied by using computer simulations
(both Monte Carlo and Molecular Dynamics) or solving some approximate integral
equation (IE) for correlation functions, derived from the OZ equation.

The simplest model for electrolyte solutions is the so-called
\textit{primitive model} (PM), in which the ions of species $i$ are depicted
as charged hard spheres, with hard sphere (HS) diameter $\sigma_{i}$ and
charge $z_{i}e$ ($e$ is the elementary charge, and $z_{i\text{ }}$the
electrovalence), embedded in a solvent represented as a continuum with
dielectric constant $\varepsilon$. The corresponding interaction potential
reads as%
\begin{equation}
\label{long:eq1}
\phi_{ij}^{\mathrm{PM}}(r)=\left\{
\begin{array}
[c]{cc}%
+\infty & r<\sigma_{ij}\equiv(\sigma_{i}+\sigma_{j})/2\\
z_{i}z_{j}e^{2}/\left(  \varepsilon r\right)  & r\geq\sigma_{ij}%
\end{array}
\right.
\end{equation}
and is long-ranged, due to its Coulombic tail added to the HS repulsion term
($i$ and $j$ are species indexes).

The OZ integral equation (\ref{simple:eq1}) can then be extended to multicomponent mixtures,

\begin{equation}
\label{long:eq2}
h_{ij}(r)=c_{ij}(r)+\sum_{l}\rho_{l}\int d\mathbf{r}^{\prime}\ h_{il}%
(r^{\prime})c_{lj}(\left\vert \mathbf{r-r}^{\prime}\right\vert )
\end{equation}
and defines the direct correlation functions $c_{ij}(r)$ in terms of the total
correlation functions $h_{ij}(r)\equiv g_{ij}(r)-1$, with $\rho_{l}$ and
$g_{ij}(r)$ being respectively the number density of species $l$ and the
radial distribution functions (RDF). As in simple fluids, the presence of the HS exclusion between
ionic spherical volumes leads to the exact core condition
\begin{equation}
\label{long:eq3}
h_{ij}(r)=-1\text{ \ \ \ \ when \ }r<\sigma_{ij}\text{.\ }%
\end{equation}
Once more, in order to determine both $c_{ij}(r)$ and $h_{ij}(r)$, a second approximate
relationship among $h_{ij}(r)$, $c_{ij}(r)$ and $\phi_{ij}(r)$ (a closure) is then required.

The simplest closure useful for ionic fluids is the extension to mixtures of the MSA closure
defined in Eq.(\ref{simple:eq3}). Although the
HNC closure is typically more accurate for ionic fluids, it can be solved only numerically. The main
advantage of the MSA is that the OZ-MSA equations can instead be solved analytically.

In the general case of a mixture of charged spheres with arbitrary HS
diameters, the analytical MSA solution was obtained by Blum long time ago
(Blum, 1975). Such a theory yields both pair correlation functions and
thermodynamics in terms of a single parameter $2\Gamma$, which reduces to
Debye's inverse length $\kappa_{\mathrm{D}}$ for dilute ionic solutions. In
the MSA the thermodynamic properties are formally similar to those of the
Debye-H\"{u}ckel (DH) theory, but with $\kappa_{\mathrm{D}}$ replaced by
$2\Gamma$ (in the MSA theory, the excluded volume effects are more accurately
taken into account). In particular, the result for the mean activity
coefficient is%
\begin{equation}
\label{long:eq5}
\ln\gamma_{\pm}=-\frac{\beta e^{2}}{\varepsilon}\ \Gamma\sum_{i}\frac
{x_{i}z_{i}^{2}}{1+\sigma_{i}\Gamma}%
\end{equation}
where $x_{i}$ is the molar fraction of species $i$. The MSA expression for
$\ln\gamma_{\pm}$ was used to fit real activity coefficients of some simple
electrolytes (Triolo \textit{et al.}, 1977, 1978). Such experimental data were
also reproduced by using more refined potentials, different from the PM one.
Among these,  a model which assumes that
solvated ions behave as rigid polarizable charged spheres (Ciccariello,
Gazzillo, and Dejak, 1982) and another one where the granular structure of the
solvent can also be included to some extent (Ciccariello and Gazzillo, 1982, 1985).

Within the PM potential for ionic fluids, results comparable with the HNC ones can also be obtained by simpler methods.
The first one takes into account the first-order perturbative correction in
the renormalized cluster expansion of the RDF, while the second approach is
based upon an approximation to the Helmholtz free energy coupled with a
variational optimization procedure (Ciccariello and Gazzillo, 1983).
Remarkably, the DH approximation has been shown to provide consistent energy and virial
routes for any potential in any dimensionality (Santos \textit{et al}, 2009).

Unfortunately, the MSA and all the above-mentioned theories work well only
for ions with very small values of the valence $\left\vert z_{i}\right\vert $.
In real solutions of \textit{mesoscopic}
particles -- with sizes within the range $10-10^{4\text{ }}$\AA \ -- such as
colloids, micelles, and globular proteins, the main difficulties are due to
the possible presence of high electric charges and large charge-asymmetries,
as well as to the large difference between solute and solvent molecular sizes, thus hampering the utility of these studies within this
new context. On
the other hand, quite commonly the presence of both the solvent and other
electrolytes in the solution can produce a significant screning of all
long-ranged electrostatic interactions. Furthermore, due to the large size of
the mesoscopic (or macroscopic) suspended molecules, most of the forces acting
among them are usually short-ranged, with ranges of only a small
fraction of the molecular diameters. Typical examples for nonionic
interactions in such fluids are van der
Waals attractions, hard-sphere-depletion forces, repulsions among polymer-coated colloids, and
solvation effects, including, in particular, hydrophobic bonding and
attractions between reverse micelles of water-in-oil microemulsions (Gazzillo
\textit{et al.}, 2006a).

In the next section, we then review some results obtained by our group
for models of mesoscopic fluids with only \textit{short-ranged} --
both ionic and nonionic -- interactions, illustrating also some of their
possible biological applications.

%\bigskip
%%%%%%%%%%%%%%%%%%%%%%%%%%%%%%%%%%%%%%%%%%%%%%%%%%%%%%%%%%%%%%%%%%%%%%%%%
\section{Short-ranged interactions in fluid systems}
\label{sec:short}
%%%%%%%%%%%%%%%%%%%%%%%%%%%%%%%%%%%%%%%%%%%%%%%%%%%%%%%%%%%%%%%%%%%%%%%%%
From the experimental point of view, several biophysical techniques can be
employed for obtaining quantitative data on protein-protein interactions in
solution. However, x-ray or neutron small-angle scattering (SAS) stands out
for its reliability for studying the whole structure of the solutions and under very
different experimental conditions (pH, ionic strength, temperature, etc.).

The SAS intensity divided by the
average form factor yields the measurable structure factor (Hansen and McDonald 2006)  
\begin{equation}
\label{short:eq1}
S_{\mathrm{M}}(q)=\sum_{i,j}\left(  x_{i}x_{j}\right)  ^{1/2}w_{i}%
(q)w_{j}(q)\ S_{ij}(q),%
\end{equation}
with%
\begin{equation}
w_{i}(q)=\frac{F_{i}(q)}{\sqrt{\sum_{k}x_{k}F_{h}^{2}(q)}},
\end{equation}
and where $q$ is the magnitude of the scattering vector, $F_{i}(q)$ the angular
average of the form factor of species $i$, and the Ashcroft-Langreth partial
structure factors (for spherically-symmetric intermolecular potentials) are
given by%

\begin{equation}
\label{short:eq2}
S_{ij}(q)=\delta_{ij}+4\pi\left(  \rho_{i}\rho_{j}\right)  ^{1/2}\int
_{0}^{\infty}dr \, r^{2}h_{ij}(r)\frac{\sin\left(  qr\right)  }{qr} %
\end{equation}
in terms of the three-dimensional Fourier transform of $h_{ij}(r)$
($\delta_{ij}$ being the Kronecker delta).

Theoretical protocols envisage the inverse procedure of the choice of an appropriate potential
model to compute correlation functions $h_{ij}(r)$, that can then be used
to fit the experimental data with the theoretical scattering intensity via equations (\ref{short:eq1}) and
(\ref{short:eq2}). 
%%%%%%%%%%%%%%%%%%%%%%%%%%%%%%%%%%%%%%%%%%%%%%%%%%%%%%%%%%%%%%%%%%%%%%%%%
\subsection{Screened ionic interactions: Yukawa model}
\label{subsec:screened}
%%%%%%%%%%%%%%%%%%%%%%%%%%%%%%%%%%%%%%%%%%%%%%%%%%%%%%%%%%%%%%%%%%%%%%%%%
In the case of charged protein molecules interacting via screened Coulomb
forces, a simple model can be obtained by adding to a HS repulsion term a tail
(for $r\geq\sigma_{ij}$) given by the \textit{Yukawa} potential%
\begin{equation}
\label{screened:eq1}
\phi_{ij}^{\text{Y}}(r)=\frac{z_{i}z_{j}e^{2}}{\varepsilon(1+\kappa_{D}%
\sigma_{i}/2)(1+\kappa_{D}\sigma_{j}/2)}\ \frac{\exp[-\kappa_{D}(r-\sigma
_{ij})]}{r}, %
\end{equation}
which represents an \textit{effective} interaction between two
\textit{isolated} macroions \textit{in a sea of microions}. The functional
form is the same as in the DH theory of electrolytes, but the coupling
coefficients are of the well-known DLVO (Derjaguin-Landau-Vervey-Overbeek)
type (Vervey and Overbeek 1948). Here, the inverse Debye screening length $\kappa_{D}$ is
due only to microions, i.e.
\begin{equation}
\label{screened:eq2}
\kappa_{D}=\sqrt{\frac{8\pi\beta e^{2}}{\varepsilon}\frac{N_{A}}{1000}\left(
I_{c}+I_{s}\right)  }, %
\end{equation}
where $N_{A}$ is the Avogadro number, and $I_{c}=(1/2)c_{c}z_{c}^{2}$ denotes
the ionic strength of the counterions originated from the ionization of the
protein macromolecules ($c_{c}$ being the molar concentration of these
counterions), while $I_{s}=(1/2)\sum_{i}c_{i}^{\mathrm{micro}}\left(
z_{i}^{\mathrm{micro}}\right)  ^{2}$ is the ionic strength of all microions
(cations and anions) generated by added salts. Clearly, $\kappa_{D}^{-1}$ is
temperature-dependent and indicates the range of the screened Coulomb
interactions: $\kappa_{D}\rightarrow0$ corresponds to pure Coulomb potentials,
while $\kappa_{D}\rightarrow\infty$ yields the opposite HS limit.

For a general HS-Yukawa (HSY) model %
\begin{equation}
\label{screened:eq3}
\phi_{ij}^{\mathrm{HSY}}(r)=\left\{
\begin{array}
[c]{cc}%
+\infty & r<\sigma_{ij}\\
A_{i}A_{j}\exp[-z(r-\sigma_{ij})]/r\text{ \ \ } & r\geq\sigma_{ij}%
\end{array}
\right.  ,
\end{equation}
Blum \textit{et al. }(Blum and H\o ye, 1978; Ginoza, 1990) found an analytical
solution of the OZ equations, again within the MSA closure. Nevertheless,
under common experimental regimes corresponding to low density and strong
electrostatic repulsions (weak screening), the MSA displays a serious
drawback, since RDFs may assume unphysical negative values close to the
contact distance $\sigma_{ij}$, for particles $i$ and $j$ which repel each
other. To overcome this shortcoming for repulsive Yukawa models, one has to
resort to different closures. In general then, only numerical solutions are
feasible, and thus IE algorithms can hardly be included into best-fit programs
for the analysis of SAS results. The use of analytical solutions, or simple
approximations requiring only a minor computational effort, is clearly much
more advantageous when fitting experimental data.

The HSY model was employed for instance in the interpretation of
Small-Angle-X-ray Scattering (SAXS) measurements on structural properties of a
particular globular protein, the $\beta$-lactoglobulin ($\beta$LG), in acidic
solutions (pH = 2.3) at several values of ionic strength in the range 7-507 mM
(Baldini \textit{et al} 1999, Spinozzi \textit{et al} 2002). In particular, this protein clearly exhibits a monomer-dimer
equilibrium, affected by the ionic strength of the solution. A
\textquotedblleft two-component macroion model\textquotedblright, with
\textit{repulsive} HSY interactions between two species with charges of the
same sign (mimicking both monomers and dimers of $\beta$LG), was shown to
provide a remarkable good description of the above-mentioned SAXS data  (Spinozzi \textit{et al} 2002).
The crucial feature of that study was the proposal of a relatively
simple approximation to the RDFs, requiring a little computational effort and
thus suitable for best-fit programs, and applicable to other spherically-symmetric potentials. We review here the main idea.

The RDF can be expressed exactly in terms of the potential of mean
force $W_{ij}(r)$ as%
\begin{equation}
\label{screened:eq4}
g_{ij}(r)=\exp\left[  -\beta W_{ij}(r)\right]  .
\end{equation}
In the zero-density limit $\rho\rightarrow0$, $W_{ij}(r)$ reduces to the
potential and $g_{ij}\left(  r\right)  $ thus becomes the Boltzmann factor $g_{ij}\left(  r\right)  =\exp\left[  -\beta\phi_{ij}\left(  r\right)
\right]  $. This zeroth-order (W0) approximation was frequently used in the
analysis of experimental scattering data, since it avoids the problem of
solving the OZ equations, but is largely inaccurate except, perhaps, at
extremely low densities. Its first-order
perturbative correction (W1-approximation) reads  (Spinozzi \textit{et al} 2002)
\begin{equation}
\label{screened:eq5}
g_{ij}\left(  r\right)  =\exp\left[  -\beta\phi_{ij}\left(  r\right)
+\omega_{ij}^{(1)}(r)\rho\right]  ,
\end{equation}
with $\omega_{ij}^{(1)}(r)$ given by
\begin{equation}
\label{screened:eq6}
\omega_{ij}^{(1)}(r)=\sum_{l}{x_{l}}\int\mathrm{d}\mathbf{r}^{\prime}%
~f_{il}\left(  r^{\prime}\right)  ~f_{lj}\left(  |\mathbf{r-r}^{\prime
}|\right)  .
\end{equation}
These convolution integrals of the Mayer functions%

\begin{equation}
\label{screened:eq7}
f_{ij}\left(r\right)  =\exp\left[  -\beta\phi_{ij}\left(  r\right)
\right]  -1
\end{equation}
can be easily calculated in bipolar coordinates. It is worth
noting that this W1-expression for $g_{ij}\left(  r\right)  $ is never
negative, thus overcoming the major drawback of MSA (the monomer effective
charge resulting from the fit falls near $20e$).
As expected, the W1-approximation largely improves the fit with respect to the
W0-one (Spinozzi\textit{ et al.}, 2002).

A test of W1 against MC simulations confirmed that the first-order approximation W1 is valid for the
considered system even at strong Coulomb coupling, provided that the screening
is not too weak, i.e. for Debye length smaller than monomer radius (Giacometti
et al., 2005). 

\begin{figure}[htpb]
\begin{center}
\vskip0.5cm
\includegraphics[width=8cm]{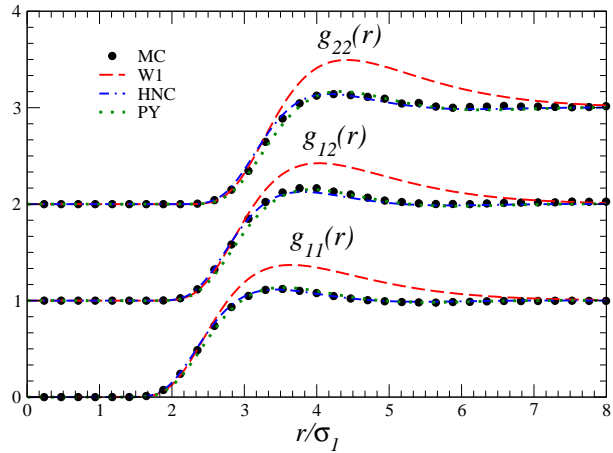}
\caption{(Color online) Plot of the $g_{11}(r)$, $g_{12}(r)$, $g_{22}(r)$ for a Yukawa potential. Points refer to MC simulations
carried out under relevant experimental conditions \cite{Spinozzi02}, and compared with integral equations
(PY dotted line, HNC dash-dotted line) and with the first-order W1 expansion (dashed line).
\label{fig:fig1}}
\end{center}
\end{figure}

An example of these results (Giacometti
et al., 2005) is depicted in Figure \ref{fig:fig1} where the three components $g_{ij}(r)$
are computed from numerical simulations under the experimetal conditions studied in
(Spinozzi et al 2002), and compared with integral equation theory
(PY and HNC) and W1 approximation.

%Fig 1 %
%%%%%%%%%%%%%%%%%%%%%%%%%%%%%%%%%%%%%%%%%%%%%%%%%%%%%%%%%%%%%%%%%%%%%%%%%%%%%%%%%%
\subsection{Nonionic surface interactions: adhesive hard sphere model}
\label{subsec:nonionic}
%%%%%%%%%%%%%%%%%%%%%%%%%%%%%%%%%%%%%%%%%%%%%%%%%%%%%%%%%%%%%%%%%%%%%%%%%%%%%%%%%%
In colloidal suspensions of \textit{neutral} mesoscopic molecules,
\textit{attractive} forces working at very short distances are quite common.
These interactions may stem from van der Waals (or dispersion) attractions
between induced dipoles, depletion forces in colloid-polymer mixtures,
solvation effects, hydrophobic bonding, and so on (Gazzillo \textit{et al.}, 2006a).

The simplest model mimicking these very short-ranged
attractive interactions is the \textquotedblleft
adhesive\textquotedblright\ or \textquotedblleft sticky\textquotedblright%
\ hard spheres (SHS), which adds to a HS interparticle repulsion an infinitely
strong attraction when the molecular surfaces come to contact. Baxter proposed
the one-component version of this model, solved the OZ equation with the
PY closure, and found that such a system presents a liquid-gas
phase transition (Baxter, 1968). The adhesive surface contribution is defined
by a particular limiting case of a square-well (SW) tail, in which the depth
goes to infinity as the width goes to zero, in such a way that the
contribution to the second virial coefficient $B_{2}$ remains finite and different from zero
(Baxter's \textquotedblleft sticky limit\textquotedblright). More
explicitly, in the one-component case, one starts from the particular SW
potential %

\begin{equation}
\label{nonionic:eq1}
\phi^{\mathrm{Baxter}\text{ }\mathrm{SW}}(r)=\left\{
\begin{array}
[c]{cc}%
+\infty & r<\sigma\\
-\epsilon\equiv-\beta^{-1}\ln\left(  1+\frac{t}{\Delta}\right)  \text{
\ \ \ \ \ } & \sigma\leq r<R\equiv\left(  1+\Delta\right)  \sigma\\
0 & r\geq R
\end{array}
\right.
\end{equation}
where $\epsilon>0$ and $w\equiv R-\sigma=\Delta\cdot\sigma$ denote the depth
and the width of the SW, respectively, while $t$ is simply related to Baxter's
original parameter $\tau$ by
\begin{equation}
\label{nonionic:eq2}
t=\frac{1}{12\tau}%
\end{equation}
Here, $t$ measures the strength of surface adhesiveness or \textquotedblleft
stickiness\textquotedblright\ between particles, and is an unspecified
\textit{decreasing} function of $T.$ In fact, as $T\rightarrow\infty$ one must
get $\tau\rightarrow\infty$, in order to recover the correct limit
corresponding to a HS fluid.

For this SW potential, the Mayer function and the second virial coefficient
become
\begin{equation}
\label{nonionic:eq3}
f^{\ \mathrm{Baxter}\text{ }\mathrm{SW}}(r)=\left\{
\begin{array}
[c]{ll}%
-1 & \qquad0<r<\sigma\\
t\ \Delta^{-1} & \qquad\sigma\leq r\leq R\\
0 & \qquad r\geq R
\end{array}
\right.  ,
\end{equation}

\begin{equation}
\label{nonionic:eq4}
B_{2}^{\ \mathrm{Baxter}\text{ }\mathrm{SW}}=-2\pi\int_{0}^{\infty}%
dr\ r^{2}f(r)=\frac{2\pi}{3}\sigma^{3}\left\{  1-\frac{t}{\Delta}\left[
\left(  1+\Delta\right)  ^{3}-1\right]  \right\}  ,
\end{equation}
respectively. Taking the \textquotedblleft sticky limit\textquotedblright%
\ $\Delta\rightarrow0$, the Mayer function becomes%
\begin{equation}
\label{nonionic:eq5}
f^{\ \mathrm{SHS}}(r)=\left[  \Theta\left(  r-\sigma\right)  -1\right]
+t\ \sigma\ \delta\left(  r-\sigma\right) %
\end{equation}
with $\Theta(x)$ being the Heaviside function ($=0$ when $x<0$, and $=1$ when
$x>0$) and $\delta(x)$ the Dirac delta function, while the SHS second virial
coefficient is simply
\begin{equation}
\label{nonionic:eq6}
B_{2}^{\mathrm{SHS}}=B_{2}^{\mathrm{HS}}\left(  1-3t\right)  =\frac{2\pi}%
{3}\sigma^{3}\left(  1-\frac{1}{4\tau}\right)  . %
\end{equation}
Here $B_{2}^{\mathrm{HS}}=2 \pi \sigma^3/3$ is the second virial coefficient of hard-spheres.
The surface adhesion corresponds to the term $t\ \sigma\ \delta\left(
r-\sigma\right)  $, where the Dirac $\delta$ ensures that the interaction
occurs only at contact, i.e. when $r=\sigma.$

Although it is known that SHSs of equal diameter, when treated exactly rather
than in the PY approximation, are not thermodynamically stable (Stell 1991), the SHS-PY solution, both for the one-component case and
for mixtures, has nevertheless received a continuously growing interest 
in the last decades as a useful model
for colloidal suspensions, micelles, microemulsions and globular protein
solutions, mainly because of its capability to exhibit a gas-liquid phase transition.

In addition to Baxter's model and its PY solution (that hereafter will be referred to as SHS1 model and
SHS1-PY solution, respectively), more than two decades ago some authors
proposed an alternative definition of adhesive-hard-sphere model, starting
from the particular HSY potential
\begin{equation}
\label{nonionic:eq7}
\beta\phi^{\mathrm{HSY}}(r)=\left\{
\begin{array}
[c]{cc}%
+\infty & r<\sigma\\
-zK_{0}\exp[-z(r-\sigma)]/r\text{ \ \ } & r\geq\sigma
\end{array}
\right. %
\end{equation}
and performing a \textquotedblleft sticky limit\textquotedblright, which in
this case amounts to taking $z\rightarrow+\infty$. The resulting model (denoted as SHS2) admits an analytic solution within the MSA closure, and
this SHS2-MSA solution is somehow simpler than the original SHS1-PY one.
Consequently, several authors utilized the SHS2-MSA solution, under the --
implicit or explicit -- assumption that both the SHS1 and SHS2 model were
different but equivalent representations of a \textit{unique} SHS potential.
Unfortunately, this turns out not to be the case, and the
usage of both the above-mentioned models has generated some confusion in the past. The problem is not simply
that the SHS2-MSA is a different solution, but -- more drastically -- that, unlike SHS1, the SHS2 model \textit{itself} is not well defined
(Gazzillo and Giacometti, 2003a), as the \textit{exact} second virial coefficient of the HSY potential \textit{diverges} in the sticky limit $z\rightarrow+\infty$.

Within the MSA, this pathology of the SHS2 model is masked by the closure, at least 
in the SHS2-MSA compressibility equation of state (EOS), whereas both the virial and the
energy MSA pressures of the HSY fluid diverge in the sticky limit (Gazzillo and Giacometti, 2003a).

Gazzillo and Giacometti (2003a) introduced a simple alternative model (SHS3), in such a way that the exact $B_{2}$
remains finite in the sticky limit, and yet the OZ equations can still be
analytically solved within a new closure (the modified MSA, or mMSA).
Rather interestingly, the SHS3-mMSA solution has exactly the same formal expression as
the SHS2-MSA one and this remarkable equivalence allows to recover all
previous results for structure factors and compressibility EOS based upon the
SHS2-MSA solution, after an appropriate re-interpretation.
%%%%%%%%%%%%%%%%%%%%%%%%%%%%%%%%%%%%%%%%%%%%%%%%%%%%%%%%%%%%%%%%%%%%%%%%%%%%%%%
\section{Polydispersity}
\label{sec:poldispersity}
%%%%%%%%%%%%%%%%%%%%%%%%%%%%%%%%%%%%%%%%%%%%%%%%%%%%%%%%%%%%%%%%%%%%%%%%%%%%%%%
Polydispersity is a rather common feature in both ionic and nonionic
suspensions of colloids and micelles. It means that
mesoscopic suspended particles of a same chemical species are not necessarily
identical, but some molecular properties (size, charge, etc.) may exhibit a
discrete or continuous distribution of values. Thus, even when all
macroparticles belong to a unique chemical species, a polydisperse fluid is
practically a multicomponent mixture, with a very large number $p$ of
components -- of order $10^{1}\div10^{3}$ or more (\textit{discrete
polydispersity}) -- or with $p\rightarrow\infty$ (\textit{continuous
polydispersity}).

In a fluid of spherical molecules with polydispersity only in size, each
component of the mixture is characterized by a different diameter. In the case
of discrete polydispersity, one can specify a large set of possible diameters
$\left\{  \sigma_{i}\right\}  $ (with $i=1,\ldots,p$) and the molar fractions
$\left\{  x_{i}\right\}  $ of the corresponding components. In the continuous
case, all diameters $\sigma\geq0$ must be included and the discrete molar
fractions must be replaced by a continuous distribution, according to the
following substitution rules%
\begin{equation}%
\label{polydispersity:eq1}
\begin{array}
[c]{c}%
x_{i}\rightarrow dx=f\left(  \sigma\right)  d\sigma\\
\sum_{i}x_{i}\ \ldots\rightarrow\int d\sigma\ f\left(  \sigma\right)  \ \ldots
\end{array}
\end{equation}
where $f\left(  \sigma\right)  d\sigma$ is the probability of finding a
particle with diameter in the range $\left(  \sigma,\sigma+d\sigma\right)  $,
and the distribution function $f\left(  \sigma\right)  $ (\textit{molar
fraction density function}) is normalized.

\begin{figure}[htpb]
\begin{center}
\vskip0.5cm
\includegraphics[width=8cm]{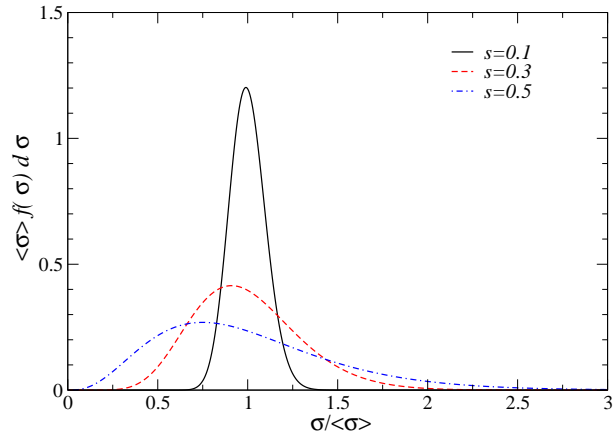}
\caption{(Color online) Plot of the Schultz distribution for different values of the polydispersity
parameter $s$.
\label{fig:fig2}}
\end{center}
\end{figure}

A common, but not unique, choice for $f\left(  \sigma\right)  $ is the Schulz
(or gamma) distribution (see Fig. \ref{fig:fig2})
\begin{equation}
f\left(  \sigma\right)  =\frac{b^{a}}{\Gamma\left(  a\right)  }\ \sigma
^{a-1}e^{-b\sigma}\text{ \ \ (}a>1\text{),}%
\end{equation}
where $\Gamma$ is the gamma function, and the two parameters $a$ and $b$ can
be expressed as $a=1/s^{2}$ and $b=a/\left\langle \sigma\right\rangle $, in
terms of the mean value $\left\langle \sigma\right\rangle $ and the relative
standard deviation $s\equiv\sqrt{\left\langle \sigma^{2}\right\rangle
-\left\langle \sigma\right\rangle ^{2}}/\left\langle \sigma\right\rangle $.
The dispersion parameter $s$, varying in the range $(0,1)$, measures the
degree of polydispersity (typical experimental values of $s$ lie in the range
$0\div0.3$). When $s\rightarrow0$ the Schulz distribution reduces to a Dirac
$\delta$ centered at $\left\langle \sigma\right\rangle $ (monodisperse limit), 
for small $s$ values $f\left(  \sigma\right)$ is very similar to a
Gaussian distribution (without its drawback of unphysically negative
diameters) and finally, for $s$ closer to $1$, $f\left(  \sigma\right)  $ becomes
asymmetric, with a long tail at large diameters.
%Fig 2 %
In the presence of polydispersity the interpretation of experimental
measurements is a hard task, as polydispersity can
significantly affect the microscopic structure of mesoscopic fluids, and must
thus be taken into account in the analysis of their SAS or thermodynamic data.
Unfortunately, polydispersity and large size and/or charge asymmetries
represent a serious challenge to the available theoretical tools, such as MC
or molecular dynamics simulations as well as to integral equations.

%%%%%%%%%%%%%%%%%%%%%%%%%%%%%%%%%%%%%%%%%%%%%%%%%%%%%%%%%%%%%%%%%%%%%%%%%%%%
\subsection{Effects on the structure factor}
\label{subsec:effects_sk}
%%%%%%%%%%%%%%%%%%%%%%%%%%%%%%%%%%%%%%%%%%%%%%%%%%%%%%%%%%%%%%%%%%%%%%%%%%%
If one is interested in the static structural properties of polydisperse
fluids, then the measurable structure factor $S_{\mathrm{M}}(q)$ for x-ray,
neutron or light SAS is usually considered, as already remarked.

Blum pionieered the first theoretical attempts in this context, by obtaining
a closed analytical expression 
for the scattering intensity from a HS mixture, valid for
\textit{any} number $p$ of components and even in the case of continuous size
polydispersity (Vrij, 1978; Blum and Stell, 1979).

Building upon these fundamental contributions, an extension of these calculations
to ionic fluids was subsequently proposed (Gazzillo \textit{et al.},
1997). Approximate scattering functions for polydisperse ionic colloidal
fluids were also obtained by a corresponding-states approach, originally
proposed for non-ionic polydisperse mixtures (Gazzillo \textit{et al.},
1999a, 1999b; Carsughi \textit{et al.}, 2000). For simplicity, the charge
polydispersity of the macroions was assumed to be fully correlated to their
size polydispersity, as the charge (or valence) of each macroion was
set to be proportional to its surface area, i.e.%

\begin{equation}
\label{effects_sk:eq1}
z\left(  \sigma\right)  =z_{\left\langle \sigma\right\rangle }\left(
\frac{\sigma}{\left\langle \sigma\right\rangle }\right)  ^{2},
\end{equation}
where $z_{\left\langle \sigma\right\rangle }$ denotes the valence of the
macroions having diameter $\left\langle \sigma\right\rangle $.

Ginoza and Yasutomi (1998a, 1998b) obtained analytical expressions for the structure factor
of colloidal dispersions with screened Coulomb interactions, in the presence
of both size and charge polydispersities.
They utilized the MSA solution for multicomponent HSY fluids, and analyzed
experimental data for a salt-free acqueous suspension of highly charged
polystyrene spheres ($z\approx273e$). These authors also studied the equation
of state and other thermodynamic properties of polydisperse colloidal
suspensions, again with the HSY model (Ginoza and Yasutomi, 1997, 1998c).

Additional studies by our group unraveled the structure of polydisperse sticky hard
spheres in some details (Gazzillo and Giacometti, 2000, 2002, 2003b, 2004).

\begin{figure}[htpb]
\begin{center}
\vskip0.5cm
\includegraphics[width=8cm]{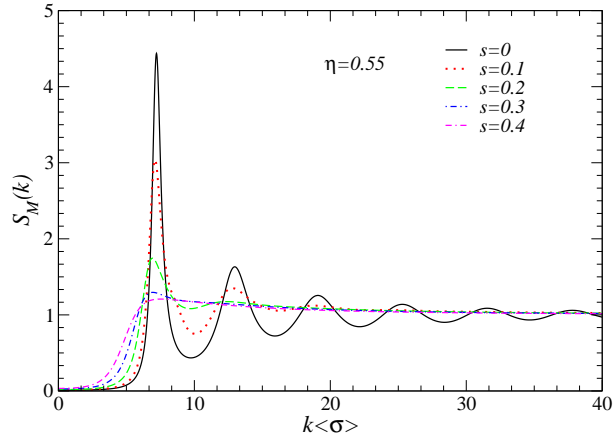}
\caption{(Color online) Plot of the effective structure factor $S_M(k)$ as a function of $k \langle \sigma \rangle$ for hard sphere potential
and different polydispersity degrees.
\label{fig:fig3}}
\end{center}
\end{figure}

From a general qualitative point of view, the effect of polydispersity on the
structure factor of a mixture is threefold, both for ionic and nonionic
fluids. As $s$ is increased at fixed packing fraction $\eta$, i) the value of
$S_{\mathrm{M}}(q=0)$ is increased, since highly dispersed particles can be
more efficiently packed; ii) the first peak is lowered, broadened and shifted
to smaller $q$ values, corresponding to a larger distance between
nearest-neighbour molecules; iii) the oscillations on the tail of the curves,
and thus the range of microscopic ordering, are greatly reduced, as a
consequence of the destructive interference stemming from the several length
scales involved (see Figure \ref{fig:fig3}).
%Fig 3 %
%\bigskip
%%%%%%%%%%%%%%%%%%%%%%%%%%%%%%%%%%%%%%%%%%%%%%%%%%%%%%%%%%%%%%%%%%%%%%%%%%%
\subsection{Effects on phase behaviour}
\label{subsec:effects_phase}
%%%%%%%%%%%%%%%%%%%%%%%%%%%%%%%%%%%%%%%%%%%%%%%%%%%%%%%%%%%%%%%%%%%%%%%%%%
Our group also addressed the phase behaviour of polydisperse (or even binary) SHS
fluids extensively, focusing on stability boundaries, percolation threshold
(related to flocculation) gas-liquid coexistence with its cloud and shadow
curves, and size fractionation effects (Fantoni \textit{et al.}, 2005a, 2005b,
2006; Gazzillo \textit{et al.}, 2006b). We considered several different cases
of polydispersity, so that for a complete description of the results we
refer to the original papers. As a general rule, the main findings are that
i) polydispersity inhibits instabilities, i.e. the
mixture becomes more stable with respect to concentration fluctuations; ii)
polydispersity increases the region of the phase diagram where the fluid is
not percolating, and iii) it diminishes the size of the two-phase coexistence region.

%%%%%%%%%%%%%%%%%%%%%%%%%%%%%%%%%%%%%%%%%%%%%%%%%%%%%%%%%%%%%%%%%%%%%%%%%%%%%%%
\section{Anisotropy effects}
\label{sec:ani}
%%%%%%%%%%%%%%%%%%%%%%%%%%%%%%%%%%%%%%%%%%%%%%%%%%%%%%%%%%%%%%%%%%%%%%%%%%%%%%%
%%%%%%%%%%%%%%%%%%%%%%%%%%%%%%%%%%%%%%%%%%%%%%%%%%%%%%%%%%%%%%%%%%%%%%%%%%%%%%%
\subsection{Failure of isotropic models for globular proteins}
\label{subsec:failure}
%%%%%%%%%%%%%%%%%%%%%%%%%%%%%%%%%%%%%%%%%%%%%%%%%%%%%%%%%%%%%%%%%%%%%%%%%%%%%%%
Until few years ago, it was widely accepted that the crucial problem to be addressed in the framework of
colloidal solutions was, as remarked, the effect of polydispersity. This view has however been challenged
recently, as a significant leap forward has been performed in the chemical syntesis of colloidal particles (Hong et al 2008).
Today there exist a number of well defined experimental protocols to produce a wide variety of colloidal 
particles with diverse anisotropies in shape, chemical composition, functionality, and surface make-up  (Walther and M\"uller 2000,
Pawar and Kretzschmar 2010). 

Even on the protein side, however, the use of spherically symmetric potentials to describe protein-protein
interactions at the tertiary level, has long ago been proven to be inadequate. This holds true both for numerical
simulations (such as Monte Carlo or molecular dynamic simulations) and integral equation theories. 
While difficulties in the extension of the former
to anisotropic potentials is mainly due to the enormous increase of computational time that requires the use of clever techniques
to maintain it within an affordable range, in the latter case a significant extension of the theoretical formalism
is also necessary. As it will be further elaborated below, some of these models can be also tackled with standard analytical tools,
although the success of these attempts has been so-far limited to qualitative agreements. These models and techniques are typically
adapted from those devised for molecular fluids (Gray and Gubbins 1984, Hansen and McDonald 2006). 

There are two additional pieces of information that can be given at this point. Firstly, that RHNC provides a rather precise description of both structural
and thermodynamical properties arising from typical isotropic potentials such as, for instance, the square-well (SW), as it appears from a detailed comparison
with MC simulations. This was illustrated recently by a detailed analysis that addressed specifically the derivation of the gas-liquid coexistence
curve (binodal) using integral equations (Giacometti \textit{et al} 2009a). Rather surprisingly, in spite of the large number of studies applying integral 
equation theory to SW, only two other examples of these calculations, both very recent and with different closures, are known in the literature, 
to the best of our knowledge (White and Lipson 2007, El Mendoub \textit{et al} 2008).

Secondly, none of the isotropic potentials can capture some of the essential features of globular proteins in solutions.
as illustrated by a comparison with small-angle-scattering experiments. This second point was inferred through a dedicated study
that compared experimental results stemming from small-angle-scattering to MC simulations and integral equations (PY and HNC)
for Yukawa potentials (Giacometti \textit{et al} 2005).

%%%%%%%%%%%%%%%%%%%%%%%%%%%%%%%%%%%%%%%%%%%%%%%%%%%%%%%%%%%%%%%%%%%%%%%%%%%
\subsection{Patchy and dipolar-like models}
\label{subsec:models}
%%%%%%%%%%%%%%%%%%%%%%%%%%%%%%%%%%%%%%%%%%%%%%%%%%%%%%%%%%%%%%%%%%%%%%%%%%%
We now analyse some of the theoretical models that have been recently discussed in this framework. The selection is only meant to be representative and
largely influenced by the interests of the authors in the field.

Let us focus on one of the experimental techniques, called templating, that has proven to be one of the most efficient to directly access to
the surface of a colloidal particle and modify its functionality in a controlled and heterogeneous way (Hong \textit{et al} 2008). In this paper, the authors
exploited the so-called Pickering technique, that is an efficient way of trapping colloidal particles at an emulsion interface, combined with epifluorescence
microscopy, that is a useful tool to visualize them, to obtain a set of amphiphilic colloidal particles where one of the two hemispherical surfaces is
hydrophobic and the other is charged. If the division is even ($50\%$ hydrophobic and $50\%$ charged), these spheres are a nice example of the
so-called Janus particles. When Janus particles are inserted into pure water, they tend to form isolated monomers irrespective of their
orientation, because of the strong repulsions. One the other hand, one can screen the repulsion by adding suitable salt, so that clusters with 
complex morphologies can be obtained depending on the salt concentration.

An interesting minimal model to tackle this problem was proposed by Kern and Frenkel (2003)  patterned after a similar model
introduced by Chapman, Jackson and Gubbins (Chapman \textit{et al} 1988) in the framework of molecular fluids.

The Kern-Frenkel potential reads 
\begin{eqnarray}
\label{models:eq1}
\Phi\left(12\right) &=& \phi \left(r_{12}\right) \Psi\left(\hat{\mathbf{n}}_1,
\hat{\mathbf{n}}_2,\hat{\mathbf{r}}_{12} \right),
\end{eqnarray}
where the radial part is a SW potential
\begin{equation}
\label{models:eq2}
\phi\left(r\right)= \left\{ 
\begin{array}{ccc}
\infty,    &  &   0<r< \sigma    \\ 
- \epsilon, &  &   \sigma<r< \lambda \sigma   \\ 
0,          &  &   \lambda \sigma < r \ %
\end{array}%
\right.  
\end{equation}
and the angular part is given by
\begin{equation}
\label{models:eq3}
\Psi\left(12\right)\equiv\Psi\left(\hat{\mathbf{n}}_1,
\hat{\mathbf{n}}_2,\hat{\mathbf{r}}_{12}\right)= \left\{ 
\begin{array}{ccccc}
1,    & \text{if}  &   \hat{\mathbf{n}}_1 \cdot \hat{\mathbf{r}}_{12} \ge \cos \theta_0 & \text{and} &  
-\hat{\mathbf{n}}_2 \cdot \hat{\mathbf{r}}_{12} \ge \cos \theta_0,   \\ 
0,    &  & &\text{otherwise}, & \
\end{array}%
\right.  
\end{equation}
$\theta_0$ being the angular semi-amplitude of the patch. Here $\hat{\mathbf{n}}_1(\omega_1)$ and $\hat{\mathbf{n}}_2(\omega_2)$ are unit vectors 
giving the directions of the center of the patch in spheres 1 and 2, respectively, with $\omega_1=(\theta_1,\varphi_1)$ and 
$\omega_2=(\theta_2,\varphi_2)$ being their corresponding spherical angles in an arbitrarily oriented coordinate frame. Similarly, 
$\hat{\mathbf{r}}_{12}(\Omega)$ is the unit vector of the separation $\mathbf{r}_{12}$ between the centers of mass of the two spheres, and is defined by the 
spherical angle $\Omega$. As usual, we have denoted with $\sigma$ and $(\lambda-1) \sigma$ the hard core diameter and the width of the well.

The coverage $\chi$ is defined by the fraction of attractive surface, that is 
\begin{eqnarray}
\label{models:eq5}
\chi &=& \left\langle \Psi\left(\hat{\mathbf{n}}_1,\hat{\mathbf{n}}_2,\hat{\mathbf{r}}_{12} 
\right) \right \rangle_{\omega_1 \omega_2}^{1/2} = \sin^2\left(\frac{\theta_0}{2}\right).
\end{eqnarray}
where we have defined $\langle \ldots \rangle_{\omega}=(4 \pi)^{-1}\int d \omega \ldots $ as the angular average over the solid angle $\omega$.

\begin{figure}[htpb]
\begin{center}
\vskip0.5cm
\includegraphics[width=5cm]{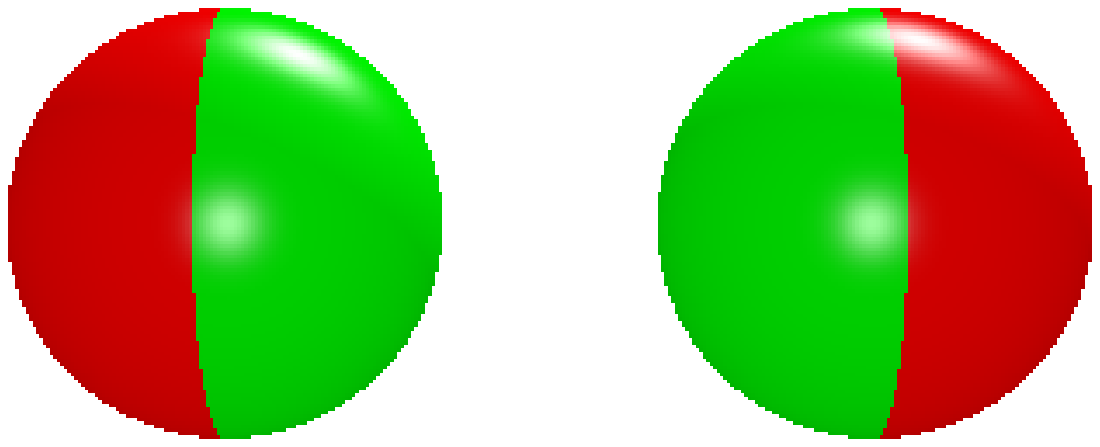} 
\includegraphics[width=5cm]{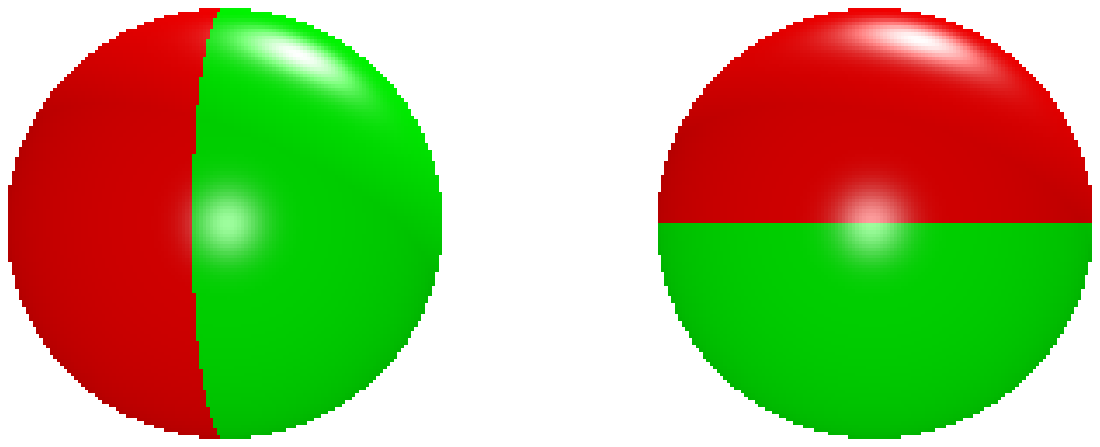} 
\includegraphics[width=5cm]{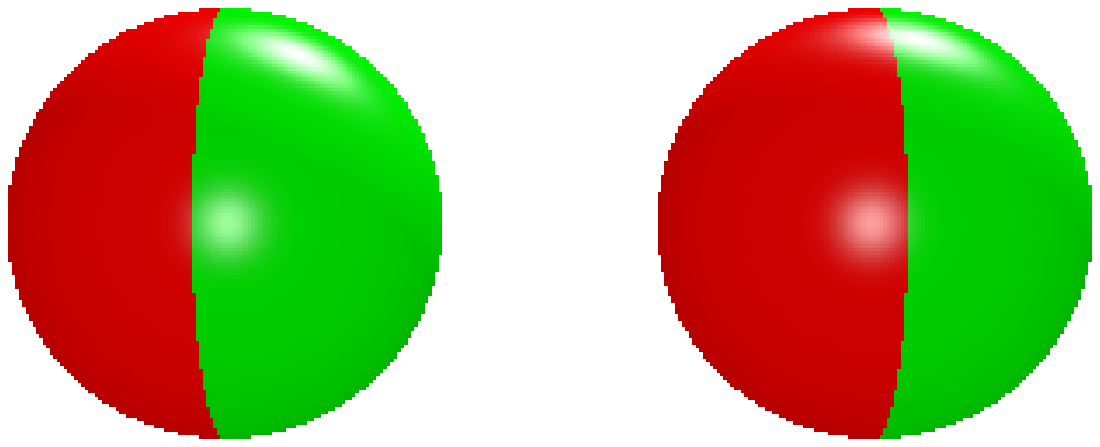} 
\caption{(Color online) Typical configurations for the Kern-Frenkel potentials under $50\%$ coverage (Janus particle $\chi=0.5$). From left to right:
head-to-head, perpendicular and head-to-tail alignment. The color coding is the following: light grey (green), attractive part; dark grey (red) repulsive part.
\label{fig:fig4}}
\end{center}
\end{figure}

The three most relevant configurations for the Janus case ($\chi=0.5$) are depicted in Figure \ref{fig:fig4}. The color coding in this figure
is that green (light gray) corresponds to the SW part, whereas red (dark grey) to the HS part. The left panel depicts the head-to-head configuration,
where the two patches are perfectly facing one another, and hence attraction is identical to the SW counterpart. The central panel, corresponds to the
perperdicular case, where there is a partial overlapping, whereas the right panel displays the head-to-tail configuration, where there
is no binding even at contact. 

%Fig 4 %
Hinging upon the SW radial counterpart, this model cannot be easily tackled by analytical means. A more convenient model for analytical
treatment was suggested by Fantoni et al (2007) and is an extension of the SHS to angular dependent interactions
where the Boltzmann factor has the following expression
\begin{eqnarray}
\label{models:eq4}
e\left(12\right)&=&\Theta(r_{12}-\sigma)+\frac{\Psi\left(12\right)}{12 \tau}\sigma \delta(r_{12}-\sigma)~.
%\label{SHS}
\end{eqnarray}
When $\Psi(12)=1$ we recover the usual Baxter SHS model (see Eq.(\ref{nonionic:eq5}))
and hence the only orientational dependence is included into the definition
of $\Psi(12)$. In very much the same way as it happens to the Kern-Frenkel potential, this reduces to the usual SW interaction
in the limit $\Psi(12)=1$, obtained whenever the fraction of surface pertaining to the SW attractive potential (i.e. the coverage)
tends to $1$, and hence $\theta_0 \to \pi$. 
The physical meaning of the above equations is rather clear. In both cases the original spherical model has been extended to include an angular dependence
that reduces the overall attraction. As we shall see, this has far reaching consequences for the resulting gas-liquid phase diagram.

Another possibility is to include continuous modulation, as opposite to reduction, of the attraction. Within the SHS framework, this can be done, for instance,
by the following Boltzmann factor (Gazzillo \textit{et al} 2008, 2009, Gazzillo 2010, 2011)
\begin{equation} 
\label{models:eq6}
e(12)=e_{\mathrm{HS}}(r)+t\ \epsilon (1,2)\ \sigma \delta
\left( r_{12}-\sigma \right) ,  
\end{equation}%

\begin{equation} 
\label{models:eq7}
\epsilon (1,2)=1+\alpha D(1,2),  
\end{equation}%
where $\alpha$ is a tunable parameter, and where we have defined the dipolar function%
\begin{equation}
\label{models:eq8}
D(12)=D\left(\hat{\mathbf{r}}_{12},\hat{\mathbf{n}}_{1},\hat{\mathbf{n}}_{2}\right)=3(\hat{\mathbf{r}}_{12}\cdot 
\hat{\mathbf{n}}_{1})(\hat{\mathbf{r}}_{12}\cdot \hat{\mathbf{n}}_{2})-\hat{\mathbf{n}}_{1}\cdot 
\hat{\mathbf{n}}_{2}  
\end{equation}%
Note that this potential can be reckoned as the analog of the dipolar potential without its radial dependence, which is here of the SHS type.
As such, the attractive part in this potential is not reduced, but only redistributed in an asymmetric way along the various directions.
%%%%%%%%%%%%%%%%%%%%%%%%%%%%%%%%%%%%%%%%%%%%%%%%%%%%%%%%%%%%%%%%%%%%%%%%%%%%%%%
\section{Analytical approaches and integral equation theories}
\label{sec:analytical}
%%%%%%%%%%%%%%%%%%%%%%%%%%%%%%%%%%%%%%%%%%%%%%%%%%%%%%%%%%%%%%%%%%%%%%%%%%%%%%%
As anticipated, the case of angular dependent potentials turns out to be far more challenging for theoretical approaches as compared with their
symmetrical counterparts, so that only very highly simplified models are actually accessible. We now review some of them into a unified perspective.
%%%%%%%%%%%%%%%%%%%%%%%%%%%%%%%%%%%%%%%%%%%%%%%%%%%%%%%%%%%%%%%%%%%%%%%%%%%%%%%
\subsection{Virial expansion}
\label{subsec:virial}
%%%%%%%%%%%%%%%%%%%%%%%%%%%%%%%%%%%%%%%%%%%%%%%%%%%%%%%%%%%%%%%%%%%%%%%%%%%%%%%
The virial expansion has the following general form 
\begin{eqnarray}
\label{virial:eq1}
\frac{\beta P}{\rho} &=& 1+ B_2\left(T\right) \rho + B_3\left(T\right) \rho^2 + \ldots
\end{eqnarray}
For the variation of the Kern-Frenkel potential, given in Eq.(\ref{models:eq4}),
the first two virial coefficients are (Fantoni \textit{et al}, 2007)
\begin{eqnarray}
\label{virial:eq2}
B_2&=& \left(\frac{\pi \sigma^3}{6}\right)\left[4-12\frac{\chi_1}{12\tau}\right]~,\\
\label{virial:eq3}
B_3&=& \left(\frac{\pi \sigma^3}{6}\right)^2\left[10-60\frac{\chi_1}{12\tau}+144\frac{\chi_2}{(12\tau)^2}-
96\frac{\chi_3}{(12\tau)^3}\right]~,
\end{eqnarray}
where $\chi_1$, $\chi_2$ and $\chi_3$ are terms accounting for the actual reduction of the coverage angle $\theta_0$ and are thus
related to it. Explicit expressions of them, as a function of $\theta_0$, can be found in Fantoni \textit{et al} (2007).

It is worth emphasizing that if one limits the expansion to the second virial coefficient,
a law of corresponding states based on the rescaling $\tau\rightarrow
\tau/\chi_1$ between the patchy and the isotropic SHS models holds true, and this could also be connected with the
SW counterpart by computing the corresponding second virial coefficient. However, this correspondence breaks down even at the level of the third 
virial coefficient, as shown in equations (\ref{virial:eq3}). 
This clearly holds true even more for the SW case, where the second virial coefficient is not linearly related 
with temperature.
 
As in spherical potentials, the most reliable approach hinges upon integral equation theories that will be discussed next.
%%%%%%%%%%%%%%%%%%%%%%%%%%%%%%%%%%%%%%%%%%%%%%%%%%%%%%%%%%%%%%%%%%%%%%%%%%%%%
\subsection{Angular-dependent Ornstein-Zernike equation and closure}
\label{subsec:OZ}
%%%%%%%%%%%%%%%%%%%%%%%%%%%%%%%%%%%%%%%%%%%%%%%%%%%%%%%%%%%%%%%%%%%%%%%%%%%%
The more general case of SW potential for the radial part cannot be tackled analytically so one has to resort to either
numerical simulations or integral equation theories. The implementation of both turns out to be not an easy task in the case of
angular dependent potentials.

Let us here focus on the latter, where the OZ equation in terms of $\gamma(12)=h(12)-c(12)$ is the generalization of Eq.(\ref{simple:eq1})
to angular dependent potentials
\begin{eqnarray}
\label{OZ:eq1}
\gamma\left(12\right) &=& \frac{\rho}{4\pi} \int d\mathbf{r}_3 d\omega_3 \left[ \gamma\left(13\right)+c\left(13\right)\right] 
c\left(32\right),
\end{eqnarray}
while the general form of a closure is again a generalization of Eq.(\ref{simple:eq2})
\begin{eqnarray}
\label{OZ:eq2}
c\left(12\right)&=& \exp \left[-\beta \Phi\left(12\right)+\gamma\left(12\right)+
B\left(12\right)\right]-1-\gamma\left(12\right).
\end{eqnarray}

The additional procedure for implementing integral equations for angular dependent potentials is constituted by an expansion 
in spherical harmonics and by rotations among different frames through Clebsh-Gordan transformations (Gray and Gubbins 1984).

The great advantage of integral equation theory with respect to numerical simulations is that a complete description of any correlation
functions can be given in any frame, within the chosen closure. This allows to uncover some nuances that are not easily seen in
numerical simulations, in view of the fact that some sort of angular averages have to be performed in order to achieve a reasonable
statistics.

%%%%%%%%%%%%%%%%%%%%%%%%%%%%%%%%%%%%%%%%%%%%%%%%%%%%%%%%%%%%%%%%%%%%%%%%%%%%%%%
\section{Effects on the thermodynamics and structural properties}
\label{sec:effects}
%%%%%%%%%%%%%%%%%%%%%%%%%%%%%%%%%%%%%%%%%%%%%%%%%%%%%%%%%%%%%%%%%%%%%%%%%%%%%%%
The Kern-Frenkel potential described in Eqs.(\ref{models:eq1})-(\ref{models:eq3}) can be regarded as an intermediate case, bracketed by a fully isotropic SW potential
(corresponding to $\theta_0=\pi$, or $\chi=1$) and a hard-sphere potential ($\theta_0=0$, or $\chi=0$). Hence this anisotropic model smoothly interpolates between
two well known and studied isotropic fluids that can then be used as benchmarks. The main effect of the anisotropy is the reduction
of the global attractive intensity along some specific directions. As a consequence, the overall energy scale is then reduced,
but this reduction occurs in a non-trivial way, and is not predictable from a corresponding-like principle (Giacometti \textit{et al} 2009, 2010), as remarked.

%%%%%%%%%%%%%%%%%%%%%%%%%%%%%%%%%%%%%%%%%%%%%%%%%%%%%%%%%%%%%%%%%%%%%%%%%%%%%%%%
\subsection{Thermodynamics and pair correlation functions}
\label{subsec:thermodynamics}
\begin{figure}[htpb]
\begin{center}
\vskip0.5cm
\includegraphics[width=8cm]{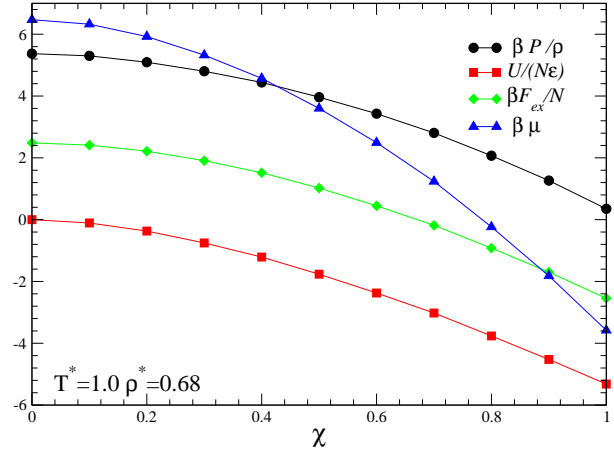}
\caption{(Color online) Plot of the thermodynamic quantities $\beta P/\rho$, $U/N\epsilon$,$\beta F_{ex}/N$, $\beta \mu$ as a function of $\chi$ for
$T^{*}=1.0$ and $\rho^{*}=0.68$ corresponding to a high density point in the gas phase, above the critical temperature. 
\label{fig:fig5}}
\end{center}
\end{figure}

%%%%%%%%%%%%%%%%%%%%%%%%%%%%%%%%%%%%%%%%%%%%%%%%%%%%%%%%%%%%%%%%%%%%%%%%%%%%%%%%
In order to have an idea of the effect of decreasing coverage $\chi$ on thermodynamics, we
plot the reduced internal energy $U/N\epsilon$, excess free energy $\beta F_{ex}/N$, chemical potential
$\beta \mu$ and virial pressure $\beta P/\rho$ for a high density state point $T^{*}=1.0$ and $\rho^{*}=0.68$ above
the critical temperature. Here $T^{*}=k_B T/\epsilon$ and $\rho^{*}=\rho \sigma^3$ are the reduced temperatures and densities, respectively.
These results were obtained by integral equation methods within the RHNC approximation
(Giacometti \textit{et al} 2009b, 2010), and are depicted in Fig.\ref{fig:fig5}.
%Fig 5 %

As one might have predicted at the outset, as $\chi$ decreases all thermodynamic quantities display a tendency
toward a decrease of the importance of attractions with respect to repulsions, that become the only interactions in the hard-sphere
limit $\chi=0$. The internal energy $U/N\epsilon$, for instance, tends to increase from negative values and vanishes as the HS limit
$\chi=0$ is reached. Similar features occur for the other thermodynamic variables. 

In these results, we note that there is a monotonic dependence on the coverage $\chi$ of all thermodynamical quantities reported in
Fig.\ref{fig:fig5}. This is due to the fact that, in their derivation, an orientational average is always present (see e.g. Eqs.(\ref{virial:eq1}) and
(\ref{virial:eq2})). This is however not mirrored by the $\chi$ dependence of the pair correlation functions $g(12)$ that, as remarked,
keeps track of the detailed orientational dependence with respect to the relative angular position $\hat{\mathbf{r}}_{12}(\Omega)$ of the two particles, 
and with respect to the relative orientation of the two unit vectors
$\hat{\mathbf{n}}_{1}(\omega)_1$ and $\hat{\mathbf{n}}_{2}(\omega_2)$ associated with the patch orientations, in addition to the relative distance $r_{12}$.

 %Fig 6 %
\begin{figure}[htpb]
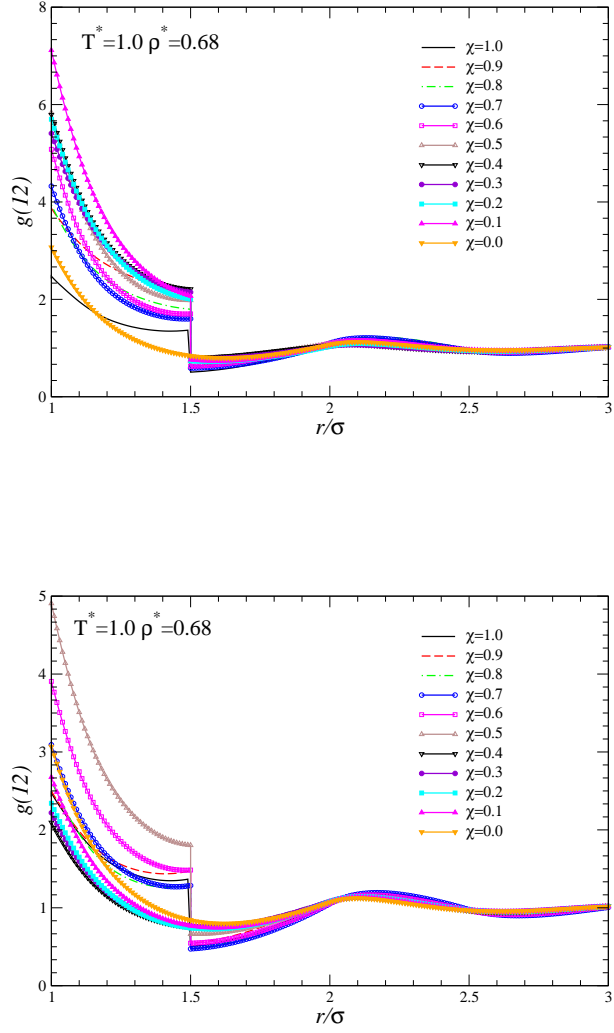

\begin{center}
\includegraphics[width=8cm]{Fig6a.eps} \\
\vskip2.0cm
\includegraphics[width=8cm]{Fig6b.eps}
\caption{(Color online) Plot of the pair correlation function $g(12)$ as a function of $r$ for
the same state point as before $T^{*}=1.0$ and $\rho^{*}=0.68$ and different coverages $\chi$. Top panel 
refers to the head-to-head configuration, bottom panel to the perpendicular counterpart (see Fig.\ref{fig:fig4}).
Different curves are distinguished by the same sequences of symbols (and colors in the on-line version) in both cases.
\label{fig:fig6}}
\end{center}
\end{figure}
This is shown in Fig.\ref{fig:fig6} where the $g(12)$ is plotted as a function of $r \equiv r_{12}$ for different coverages, 
in the molecular frame where $\hat{\mathbf{r}}_{12} = \hat{\mathbf{z}}$. 
The head-to-head configuration (top panel) represents the case where the attractive patches
are facing one another, that is $\hat{\mathbf{n}}_{2} \cdot \hat{\mathbf{n}}_{1}=1$, whereas  the other interesting case (bottom panel) the
patches are perpendicular one another, that is $\hat{\mathbf{n}}_{2} \cdot \hat{\mathbf{n}}_{1}=0$. In all cases, a herratic coverage dependence can be clearly observed. Consider for instance the head-to-head case (top panel). As $\chi$
is reduced from the isotropic SW value $\chi=1.0$ to $\chi=0.9$, the contact value at $r=\sigma^{+}$ increases, a clear indication
of the increased probability for this configuration as compared, for instance, with the perpendicular configuration (bottom panel),
where the contact value decreases. This is confirmed by a parallel increase in the discontinuity at $r=\lambda \sigma$ and mirrored 
by a corresponding reduction of the same discontinuity in the perpendicular configuration. As $\chi$ is further decreased to $\chi=0.8$,
there is a clear reduction of the discontinuity at $r=\lambda \sigma$ accompained by only a negligible variation of the contact value,
and there is no clear trend within the well width $\sigma < r < \lambda \sigma$ in both configurations, until the half coverage value
($\chi=0.5$) is reached. Interestingly, this value corresponds to the Janus limit of even philicities, and clearly plays a particular role,
as already suggested by numerical simulations of the corresponding phase diagram (Sciortino \textit{et al} 2009, 2010). 
For lower coverages, the contact values again increase for the head-to-head configurations, and decrease for the perpendicular configurations,
indicating a marked tendency to form micelles, as already observed in numerical simulations. This region of the phase diagram is inherently
interesting. As a matter of fact, at $\chi=0.5$ (Janus limit) numerical simulations show (Sciortino \textit{et al} 2009) that, at sufficiently low temperatures and
high densities, these double-face spheres tend spontaneously to aggregate into single or multiple layer clusters, to maximize all favourable contacts. 
These clusters have the attractive parts (that mimic the hydrophobic nature of the original experiments) buried inside and the hard-sphere parts
externally exposed. Upon further cooling, this micellization tendency  destabilizes more and more the liquid-gas transition that eventualy get almost
completed inhibited (Sciortino \textit{et al} 2009). This is a peculiar feature of the Janus case and it is not shared by any higher value of coverage that display
a conventional gas-liquid transition, albeit with significantly reduced critical temperatures and densities. No simulations have been able so far to probe the region
below the Janus limit (i.e. for $\chi <0.5$), due to the extremely low temperatures involved, but it is reasonable to expect again an absence of
gas-liquid transition in view of the impossibility for the system to form clusters sufficiently large to yield a percolating network. 
This case strongly contrasts with the low $\chi$ counterpart of the two-patches case, where the attractive
surface is spread on the two opposite poles of each sphere. In that case it has been shown  (Giacometti \textit{et al} 2010) that the value $\chi=0.5$ does not play
any peculiar role, and the gas-liquid transition persists for coverages as low as $30\%$. Below that value the gas-liquid transition becomes metastable
against crystalization, in very much the same way of what happens in the framework of spherical potentials of sufficiently short ranges.

\begin{figure}[htpb]
\begin{center}
\vskip0.5cm
\includegraphics[width=8cm]{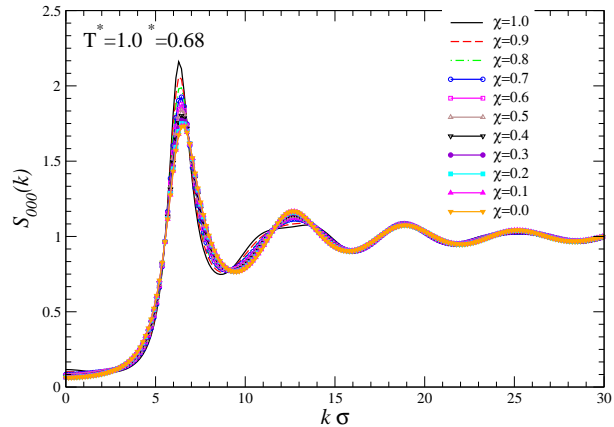} 
\caption{(Color online) Plot of the structure factor $S_{000}(k)$ for the Kern-Frenkel model as a function of $k \sigma$ for
the same state point as before $T^{*}=1.0$ and $\rho^{*}=0.68$ and different coverages $\chi$. 
Different curves are distinguished by the same sequences of symbols (and colors in the on-line version) as in Fig.\ref{fig:fig6}.
\label{fig:fig7}}
\end{center}
\end{figure}

Although the above pair correlations have been displayed in the molecular frame, where the $\hat{\mathbf{z}}$ axis is taken along the $\hat{\mathbf{r}}_{12}$
direction, integral equation theory is able to provide $g(12)$ in any arbitrary frame by performing an appropriate Clebsh-Gordan transformation 
(Gray and Gubbins 1984). 
This constitutes one of the two main advantages of this technique (the other being the limited computational cost) as compared to numerical simulations, 
where such details are virtually unaccessible. A simple way to obtain from $g(12)$ an isotropic quantity is to perform
an angular integration over all three spherical angles $\omega_1$, $\omega_2$ and $\Omega_{12}$. The result is the projection along
the spherical invariant with all three indices vanishing, that is   $g_{000}(r)$ (Giacometti \textit{et al} 2010). Likewise, the structure factor $S_{000}(k)$
can also be regarded as an isotropic angular average of anisotropic original dependence. This is also shown in Figure \ref{fig:fig7}, where again the
coverage dependence appears to be more regular, as discussed. 
%Fig 7 %
%%%%%%%%%%%%%%%%%%%%%%%%%%%%%%%%%%%%%%%%%%%%%%%%%%%%%%%%%%%%%%%%%%%%%%%%%%%%%%%%%%%%%%%%%%%%%%%%%%%%%%%%%%%%%%%%
\section{Conclusions}
\label{sec:conclusions}
In this paper we have reviewed some of the main difficulties encountered in the application of the OZ integral equation theory in the
passage from simple to complex fluids. Two novel main features appear. On the one hand, colloidal suspensions are typically 
a multi-component system, and this requires
the introduction of new techniques to cope with the difficulties arising with the mixture, some of which are an extension of those originally
devised for ionic fluids. As a representative example, we have discussed the case of mixtures formed by species of different sizes and interacting
via short-range attractive forces of the Baxter type, that combines the hard-sphere potential with an adhesive force acting only at contact.
In the limit when the number of components of the mixture becomes very large, the distribution can be replaced by a continuous description
(polydispersity), that is frequently represented by a Schultz distribution. 
We have assessed the effect of polydispersity on the structural and thermodynamical properties of the fluids, and how this can be accounted for
in the framework of integral equation theory, both analytically and numerically. This is one of the fields where the work by Prof. Lesser Blum
has made a remarkable impact.

We have additionally discussed how there exist a number of systems where surface inhomogeneities  cannot be neglected, 
even to a first approximation. These include, globular proteins and -- more recently -- patchy colloids. In both cases, the 
potentials are no longer isotropic, and orientational effects play an important role. The implementation of integral equation calculations
mirrors that devised for molecular fluids, with the important difference that the angular dependence typically
displays step-wise discontinuities, mimicking either the presence of different surface groups in globular proteins, or different functionalities
in patchy colloids. Two cases were discussed, as representative examples. In the Kern-Frenkel model, the potential is factorized into
a part depending only on the distance and an angular part, tuning the interactions in the form of an on-off form. Depending on the
form chosen for the radial dependence, the OZ integral equation theory can be tackled either analytically or purely numerically. We have discussed
both cases and displayed representative examples of the results. In the first case, the radial dependence is again considered of the Baxter type,
and this allows the use of the same approximations exploited in the isotropic counterpart, at the minimal cost of a slightly  more complicate formalism. 
A major adantage of the integral equation theory is however displayed in the cases where numerical solutions are exploited with rather accurate 
closures, such as RHNC. Here, the numerical solution typically requires only few minutes of computational effort, as compared to several weeks
of numerical simulations, and yet the obtained predictions are frequently rather accurate. Furthermore, the details of information 
that can be obtained from the analysis of pair correlation functions from integral equations
is independent of the chosen frame and is unparalleled by any numerical simulations, where some form of orientational averages are always involved in order to
have a reasonable statistics.

We believe that the potentialities of integral equation theory in complex fluids has still not be fully exploited so far, and that this methodology
may become, in the next few years, the first and most reliable tool to tackle new challenging problems, before resorting to more computational demanding
methods.  

%%%%%%%%%%%%%%%%%%%%%%%%%%%%%%%%%%%%%%%%%%%%%%%%%%%%%%%%%%%%%%%%%%%%%%%%%%%%%%%
\begin{acknowledgments}
 The support of a PRIN-COFIN 2007B58EAB grant is acknowledged. The results presented in this paper summmarizes the work
carried out by the authors in collaborations with many colleagues, including Flavio Carsughi, Raffaele Della Valle, Fred Lado, Riccardo Fantoni, 
Mark Miller, Giorgio Pastore, Peter Sollich, Andr\`es Santos, Francesco Sciortino, and Francesco Spinozzi. It is a great pleasure to dedicate this paper to Prof. 
Lesser Blum, whose work was the inspiration of many generations, including ours, working in the field.
\end{acknowledgments}
%%%%%%%%%%%%%%%%%%%%%%%%%%%%%%%%%%%%%%%%%%%%%%%%%%%%%%%%%%%%%%%%%%%%%%%%%%%%%%%

%%%%%%%%%%%%%%%%%%%%%%%%%%%%%%%%%%%%%%%%%%%%%%%%%%%%%%%%%%%%%%%%%%%%%%%%%%%%%%%
%\appendix

%\section{appendix}
%\label{app:ruelle}
%%%%%%%%%%%%%%%%%%%%%%%%%%%%%%%%%%%%%%%%%%%%%%%%%%%%%%%%%%%%%%%%%%%%%%%%%%%%%%%

%%%%%%%%%%%%%%%%%%%%%%%%%%%%%%%%%%%%%%%%%%%%%%%%%%%%%%%%%%%%%%%%%%%%%%%%%%%%%%%
% References
%%%%%%%%%%%%%%%%%%%%%%%%%%%%%%%%%%%%%%%%%%%%%%%%%%%%%%%%%%%%%%%%%%%%%%%%%%%%%%%

%\clearpage
%%%%%%%%%%%%%%%%%%%%%%%%%%%%%%%%%%%%%%%%%%%%%%%%%%%%%%%%%%%%%%%%%%%%%%%%%%%%%%%
%% Fig 1: S_{ij} (q) confronto con dati sperimentali 
%% Fig 2: Schultz (?)
%% Fig 3: S_M(q) for polydisperse HS
%%%%%%%%%%%%%%%%%%%%         FIGURES    %%%%%%%%%%%%%%%%%%%%%%%%%%%%%%%%%%%%%%%

%%%%%%%%%%%%%%%%%%%%%%%%%%%%%%%%%%%%%%%%%%%%%%%%%%%%%%%%%%%%%%%%%%%%%%%%%%%%
\end{document}